\documentclass[
aps,%
12pt,%
final,%
notitlepage,%
oneside,%
onecolumn,%
nobibnotes,%
nofootinbib,
superscriptaddress,%
noshowpacs,%
centertags]%
{revtex4}

\RequirePackage{graphicx}

\def\refitem#1{\relax}

\begin{document}
\title{Physics around the QCD (tri)critical endpoint and new challenges for femtoscopy}
\author{\firstname{K. A.} \surname{Bugaev}}
\affiliation{Bogolyubov Institute for Theoretical Physics, Kiev, Ukraine}
%
\email{bugaev@th.physik.uni-frankfurt.de}
\noaffiliation

\begin{abstract}
On the basis of exactly solvable models with the tricritical and critical endpoints I 
discuss the physical mechanism of endpoints formation which is similar to the usual liquids. It is demonstrated that the necessary condition for the transformation of the 1-st order deconfinement phase transition into the 2-nd  order phase transition at the (tri)critical endpoint is the vanishing of surface tension coefficient of large/heavy QGP bags. 
Using the novel model of the confinement phenomenon  I argue that the physical reason for the cross-over appearance at low baryonic densities is the negative value of QGP bag surface tension coefficient. This implies the existence of highly non-spherical or, probably, even fractal surfaces of large and heavy bags at and above the cross-over, which, perhaps, can be observed via some correlations. 
The model with the tricritical endpoint predicts that at the deconfinement transition line 
the volume (mass) distribution of large (heavy) QGP bags acquires 
the power law form   at the endpoint only, while in the  model with  the critical endpoint such a power law  exists  inside the mixed phase. 
The role of finite width of QGP bags is also  discussed.
\end{abstract}

\maketitle

\noindent
{\bf 1. What is missing in the statistical models of strongly interacting matter equation of state?}
Almost 25 years ago the first heavy ion experiments started the searches for a new state of matter, the quark gluon plasma (QGP). During this time there were made several nice 
discoveries, but the smoked gun of  QGP  creation  is not found yet. Despite the claims that 
the Kink \cite{Kink}, the Strangeness Horn \cite{Horn} and the Step \cite{Step}
are  the reliable signals of the onset of deconfinement \cite{Claims} it is necessary to admit that, in fact, we do not exactly know what they really signal.  
Hence, 
from the present state of   heavy ion physics   I conclude that:\\ 
(I) up to now our models are missing a few key elements which do not allow us to formulate some convincing signals of the deconfinement;\\
(II) the low energy programs at RHIC (BNL), SPS(CERN), NICA (Dubna) and FAIR (GSI)   will  be hardly successful, even, if they ``discover some irregularities'', since without theoretical back up they will convince no one;\\
(III) it is necessary to return to foundations of heavy ion physics and start a systematic work to formulate and to  account for the missing key elements. 

One of the primary goals of low energy programs mentioned above is the uncovering of
 the  (tri)critical endpoint of the QCD phase diagram.  Clearly, this task is even more hard then just the discovery of  a new state of matter and, hence, I have  serious doubts that  it can be successful with the existing theoretical background. Indeed, tens of papers discuss possible signals of the QCD critical endpoint, but neither  the physical reason of its existence nor possible  experimental consequences are under intense investigations!
Furthermore, from the very  beginning it was clear that the systems studied in collisions of heavy ions at high energies  are finite or even small, but  up to now we have just  a few  general guesses on 
how to rigorously  define the phase transition (PT) in finite systems  and how the (tri)critical endpoint is modified in such  systems. 

Therefore, here I discuss the surface tension of quark gluon bags  and their finite width which, as it is argued, are the missing two key elements. 
The role of surface tension of bags in generating the (tri)critical endpoint is discussed 
on the basis of exactly solvable statistical models with the tricritical (Model 1) \cite{QGBSTM, QGBSTMb} and critical (Model 2)\cite{QGBSTM2,QGBSTM2b} endpoints. Also the recent finding of the negative values of surface tension coefficient of bags \cite{String:09} along with 
its role for the cross-over existence are discussed. 
In addition I consider the finite width model of quark gluon bags 
\cite{QGBSTM2b,FWM,FWMb} that sets some strict   limitations on their   experimental 
studies  due  to their very  short  lifetime.  Analysis of  novel  physical phenomena associated with 
the vicinity of (tri)critical endpoint allows me to formulate the new challenges for femtoscopy 
whose study, I believe,  is  utterly necessary to qualitatively improve the present state of art. 

The work is organized as follows. In the next section the appearance of negative surface tension of QGP bags is discussed. Section 3 elucidates the role of surface tension at the (tri)critical endpoint. 
The new challenges for femtoscopy are formulated in Section 4. 


\medskip 


\noindent
{\bf 2. Surface tension of quark gluon bags.}
The role of surface tension for QGP was discussed long ago  \cite{Jaffe:1, Jaffe:2}, however, up to recently its importance for the existence of the QCD (tri)critical endpoint 
was not recognized. 
In nuclear and cluster physics 
the importance of the surface tension for the properties of endpoint 
is known from a number of  exactly solvable cluster  models with 
the 1${\rm st}$ order PT   which describe the critical endpoint properties  very well. These are the Fisher droplet model (FDM) \cite{Fisher:67,Elliott:06}
and the simplified version of  the  statistical multifragmentation  model (SMM)  \cite{simpleSMM:1}.
Both of these models are built on the assumptions that 
the difference of  the bulk  part (or the volume dependent part) of  free energy  
of two phases disappears at  phase equilibrium and that, in addition, 
the difference of the surface part (or the surface tension) of  free energy  vanishes 
at the critical point. Note that such a mechanism of the critical endpoint generation is typical 
for  ordinary liquids \cite{Fisher:67,Landau:Statmech}.
According to the  contemporary paradigm at the deconfinement region the QGP is 
a strongly interacting liquid \cite{Shuryak:sQGP}, but two major  questions are: what is the value of it surface tension  and how can we measure it?

Very recently it was possible to find out  the relation between the string tension $\sigma_{str} (T) $ of 
the unbreakable color tube of length $L$ and radius $R \ll L$ which  
connects  the static quark-antiquark pair 
 and the surface tension coefficient $\sigma_{surf} (T)$ of this tube \cite{String:09}:
\begin{equation}\label{EqBugaevIII}
\sigma_{surf} (T) = \frac{\sigma_{str} (T)}{ 2 \pi R} ~ + ~ \frac{1}{2} \, p_v (T) R \,,
\end{equation}
where $p_v (T)$ is the bulk pressure inside  the tube. 
Eq. (\ref{EqBugaevIII}) was derived by equating the free energies of confining string
and the free energy of elongated cylindrical bag \cite{String:09}. 
In fact, 
in deriving   (\ref{EqBugaevIII})  we match an ensemble of all string shapes of fixed $L$ to a mean elongated cylinder, 
which according to the original Fisher idea \cite{Fisher:67} and the results   of the 
Hills and Dales Model (HDM) \cite{Bugaev:04b,Bugaev:05a} represents a sum of all surface deformations of a given bag.
Eq. (\ref{EqBugaevIII})  allows one to determine the $T$-dependence of bag surface tension coefficient,
if $R(T)$, $\sigma_{str} (T)$ and $p_v (T)$ are known. Therefore,  Eq. (\ref{EqBugaevIII})  opens a principal possibility to determine the bags surface tension for any temperature directly from the lattice QCD simulations. 
Also it allows us to estimate  the surface tension at $T=0$. Thus, taking the typical value of the bag model pressure which is used in hadronic spectroscopy $p_v (T=0) = - (0.25)^4$ GeV$^4$ and inserting 
into (\ref{EqBugaevIII}) the lattice QCD values  $R=0.5$ fm  and $\sigma_{str} (T=0) = (0.42)^2$ GeV$^2$ 
\cite{StrTension2}, one finds  $\sigma_{surf} (T=0) = (0.2229~ {\rm GeV})^3 + 
0.5\, p_v\, R\approx (0.183~{\rm GeV})^3 \approx 157.4$ MeV fm$^{-2}$ \cite{String:09}.

The above   results allow one to  study the bag 
surface tension near the cross-over to QGP. The lattice QCD data indicate that near the deconfinement, i.e.  for $T\rightarrow T_{dec}-0$, the tube radius diverges $R \rightarrow \infty$ and  the 
string tension vanishes  as \cite{StrTension2}
\begin{equation}\label{EqBugaevXI}
\sigma_{str} (T) \, R^k \rightarrow \omega_k >0 \,,
\end{equation}
 with $k=2$. However, one can
 extend  a range for the power $k>0$ to study more general case.
The value of constant $\omega_k > 0$ is not of crucial importance   here because my main interest is in the qualitative analysis. 

Consider first the case of zero baryonic chemical potential, i.e. $\mu = 0$.  Then 
using Eqs. (\ref{EqBugaevIII}) and  (\ref{EqBugaevXI})  for $L \gg R$ 
one can calculate the total bag pressure  as
\begin{eqnarray}\label{EqBugaevXIII}
&&p_{tot} = p_v (T) {\textstyle - \frac{ \sigma_{surf} (T)}{R} \equiv \frac{ \sigma_{surf} (T)}{R} - \frac{\sigma_{str}}{\pi R^2} }  \rightarrow 
{\textstyle  
\left[ \frac{\sigma_{str}}{\omega_k} \right]^{\frac{1}{k}} \left[ \sigma_{surf} ~ - ~ \frac{\omega_k}{\pi}
\left[ \frac{\sigma_{str}}{\omega_k} \right]^{\frac{k+1}{k}} 
\right]   
} \, ,
\end{eqnarray}
which at fixed value of $\mu =0$  can be considered  as the usual equation of state of a single variable $T$. Then 
the total entropy density of the cylindrical bag is 
\begin{eqnarray}\label{EqBugaevXIV}
\hspace*{-0.5cm}&&s_{tot} = \frac{\partial ~p_{tot}}{\partial~T} \rightarrow   
{\textstyle 
 \frac{1}{k\,\sigma_{str} } \left[ \frac{\sigma_{str}}{\omega_k} \right]^{\frac{1}{k}}  
\frac{\partial ~\sigma_{str} }{\partial ~T} \, \sigma_{surf} ~ +~ 
\left[  \frac{\sigma_{str}}{\omega_k} \right]^{\frac{1}{k}}  
\frac{\partial~\sigma_{surf} }{\partial ~T} 
~ - ~ \frac{k+2}{\pi\, k} \left[ \frac{\sigma_{str}}{\omega_k} \right]^{\frac{2}{k}} \frac{\partial ~\sigma_{str} }{\partial ~T} 
}  \,. ~~~
\end{eqnarray}
The mechanical stability of the cylindrical bag means an equality of the total bag pressure Eq. (\ref{EqBugaevXIII}) to  
the outer pressure, but the thermodynamic stability requires positive value for the entropy density 
(\ref{EqBugaevXIV}).  
The   Models 1 and 2 predict  \cite{QGBSTM,QGBSTM2}  that everywhere at the 
cross-over line, except for the (tri)critical endpoint, the surface tension coefficient $\sigma_{surf}$ is 
non-zero and its derivative $\frac{\partial ~\sigma_{surf}}{\partial ~T}$ is finite at $T \rightarrow T_{dec}- 0$. 
Remembering this, from Eq.(\ref{EqBugaevXIV})  one finds  that its first term on  the right hand side is 
dominant,  since $\sigma_{str} \rightarrow 0$ and  hence 
\begin{equation}\label{EqBugaevXVII}
s_{tot} \rightarrow {\textstyle 
 \frac{1}{k\,\sigma_{str} } \left[ \frac{\sigma_{str}}{\omega_k} \right]^{\frac{1}{k}}  
\frac{\partial ~\sigma_{str} }{\partial ~T} \, \sigma_{surf}} ~> 0  \,,
\end{equation}
which requires that at $T \rightarrow T_{dec} - 0$ the surface tension coefficient must be negative 
$\sigma_{surf} (T_{dec}) < 0$, since the string melts in this limit, i.e. $\frac{\partial ~\sigma_{str} }{\partial ~T} < 0$. Actually, this result is not  surprising since the calculations 
of surface partitions for physical clusters \cite{Bugaev:04b,Bugaev:05a} and the models of quark gluon bag with 
surface tension \cite{QGBSTM,QGBSTM2} also  predict  that at low 
baryonic densities the deconfining PT is transformed into a cross-over just because the surface 
tension coefficient of large bags becomes negative in this region (see also the next section).
Eq. (\ref{EqBugaevXVII}) clearly shows that the color string model shares the possibility of negative values of bag 
surface tension coefficient available in the cross-over region. 
It is necessary to stress that  negative value of the  surface tension coefficient
 $\sigma_{surf}(T)$  for temperatures above  $T_{dec}$ does not mean anything wrong. 
Fisher argued first  \cite{Fisher:67} that 
the  surface  tension coefficient consists of energy and entropy parts which  have   opposite signs \cite{Fisher:67}. 
Therefore, $\sigma_{surf}(T) < 0 $ does not mean that the surface energy changes the sign, but it
rather  means that the surface entropy, i.e. the logarithm of the degeneracy of bags of a fixed volume, simply  exceeds their  surface energy over  $T$.  In other words, 
the number of  non-spherical bags of a fixed volume becomes so large that the Boltzmann exponent, which accounts for the energy "costs" of these bags,  cannot
suppress them anymore as rigorously was shown within the   HDM 
\cite{Bugaev:04b,Bugaev:05a}. The above results are valid for 
the baryonic chemical potential values which are smaller then that one of  the (tri)critical 
endpoint, i.e. for $\mu \le \mu_{cep}$ \cite{String:09}.

Analysis of  Eq.(\ref{EqBugaevXVII}) \cite{String:09} shows  also that for $ k >1$  the entropy density of cylindrical bag can develop a singularity
at vanishing string  tension even at finite $L$. Such a surprising  conclusion can be naturally  explained
by the appearance of fractal string surfaces \cite{String:09}.
Their appearance  at the cross-over temperature can be easily understood within 
the present  model, if one recalls that only at this   temperature the fractal  surfaces  can emerge
at no energy costs due to zero total pressure.  

From the discussion above  the {\bf  first challenge for the femtoscopy} can be  formulated as follows: 
{\it to study the emission from highly non-spherical bags with complicated and even the fractal surfaces in order to 
find the indicator which is able to distinguish the case of positive, zero and negative surface tension coefficient.}
As I argue in the next section the line   $\sigma_{surf}(T,\mu)=0$
plays an important role as the  boundary separating two different physics. 


\medskip 


\noindent
{\bf 3. The role of surface tension at the (tri)critical endpoint.}
It is well known \cite{QGBSTM,QGBSTM2,Bugaev10} that 
the most convenient way to study the  phase structure of statistical models similar to FDM and SMM  is to use the isobaric 
partition  for analyzing its rightmost singularities. The isobaric partition is 
 the Laplace transform  image  of the  grand canonical one $Z(V,T,\mu)$:
\vspace*{-0.25cm}
\begin{eqnarray}\label{ZsBugaev}
\hspace*{0.05cm}\hat{Z}(s,T,\mu) \equiv \hspace*{-.05cm}
\int\limits_0^{\infty}\hspace*{-.05cm}dV\, e^{\textstyle
-sV}\,Z(V,T,\mu) =\frac{1}{ [ s - F(s, T,\mu) ] } \,,
\end{eqnarray}

\vspace*{-0.05cm} 
\noindent 
with $F(s,T,\mu)$ containing  
the discrete $F_H$ and continuous $F_Q$ volume spectra of the bags  \cite{QGBSTM}
%
%
\begin{eqnarray}
\hspace*{-0.4cm}  F(s,T,\mu) \equiv
\hspace*{-0.1cm}F_H(s,T,\mu)+F_Q(s,T,\mu) & = &
%
\sum_{j=1}^n g_j
e^{\textstyle (\frac{\mu}{T}b_j -v_js)} \phi_j(T) +  \\
\label{FsHBugaev}
\hspace*{-0.6cm}&\hspace*{-0.75cm}+&  \hspace*{-0.45cm}
{\textstyle u(T)} \hspace*{-0.1cm} \int\limits_{V_0}^{\infty}
\hspace*{-0.1cm}\frac{dv}{v^{\tau}}\hspace*{0.1cm} e^{\textstyle [
\left(s_Q(T,\mu)-s \right)v - \Sigma(T,\mu) v^{\kappa}]  }\, .
\label{FsQ}
\end{eqnarray}
\noindent $u(T)$ and $s_Q(T,\mu)$ are continuous and, at least,
double differentiable functions of their arguments.  The particle number  density of bags with mass
$m_k$, eigen volume $v_k$, baryon charge $b_k$ and degeneracy
$g_k$ is given by 
$
\phi_k(T)   \equiv  \frac{g_k}{2\pi^2}
\int\limits_0^{\infty}\hspace*{-0.0cm}p^2dp~ \exp{\textstyle
\left[- \frac{ \sqrt{p^2~+~m_k^2}}{T} \right] } 
=  g_k \frac{m_k^2T}{2\pi^2}~{ K}_2 {\textstyle
\left( \frac{m_k}{T} \right) }
$. 

The continuous part of the  spectrum (\ref{FsQ}) generalizes 
the exponential mass spectrum introduced by Hagedorn
\cite{Hagedorn:65} and it can be  derived either within the MIT
bag model \cite{Kapusta:81} or, in more general fashion, within the  finite width model of QGP bags 
\cite{FWM,Bugaev10}. The   term $e^{-s v}$ accounts for  the hard-core
repulsion of the Van der Waals type in (\ref{FsQ}). $\Sigma(T,\mu) = \sigma_{surf} (T,\mu)/T$ 
denotes the reduced surface tension coefficient 
which has the form
\begin{eqnarray}\label{SigmaBugaev}
\Sigma(T, \mu) =
\left\{ \begin{array}{rr} \Sigma^- > 0  \,,  &\hspace*{0.1cm} T
\rightarrow T_{\Sigma} (\mu)  - 0 \,,\\
 0 \,, &\hspace*{0.1cm}  T =  T_{\Sigma} (\mu) \,, \\
\Sigma^+ < 0 \,,  &\hspace*{0.1cm} T \rightarrow T_{\Sigma} (\mu)
+ 0  \,.
\end{array} \right.
\end{eqnarray}
Such a simple surface  free   energy
parameterization in (\ref{FsQ})  is based on  the original Fisher idea
\cite{Fisher:67} which allows one to account for the surface
free energy by considering 
a mean bag of volume $v$ and surface extent
$v^{\kappa}$. 
 The power $\kappa < 1$ inherent in bag
effective surface is a constant which, in principle, may differ from
the usual FDM and SMM value $\frac{2}{3}$ \cite{QGBSTM, QGBSTM2}.
From (\ref{SigmaBugaev}) one can see  that, in contrast to FDM and SMM,   the precise
disappearance of $\Sigma(T,\mu)$ above the critical endpoint is  not required. 

The Model 1  corresponds to Fisher parameter $1 < \tau \le 2$ and continuous values of  function $\Sigma(T, \mu)$ and its first derivatives. Under these conditions and for a reasonable choice of 
other parameters the  Model 1 has the 1-st order deconfinement PT and the second order 
PT at the line $\Sigma(T, \mu) = 0$ for $\mu \ge  \mu_{cep}$ and for  temperatures larger than the temperature of the  deconfinement PT $T_{dec} (\mu)$. As once can see from   Eq.  (\ref{FsQ})  the volume distribution of large QGP bags  has the  power law $1/v^\tau$  at the 
line $\Sigma(T, \mu) = 0$ \cite{QGBSTM,QGBSTMb} for $\mu \ge  \mu_{cep}$. 
This model allows one to naturally interpret the possible states on quark gluon matter. Thus, the state in which dominates a single QGP bag of infinite size is similar to the usual liquids and, hence, it can be called the {\bf  quark gluon liquid}. It is located for $\mu >  \mu_{cep}$ and temperatures satisfying the inequalities  $ T_{dec} (\mu) \le T < T_\Sigma (\mu) $. On the other hand, the state existing at  temperatures above $T_\Sigma (\mu)$ consists of the bags of finite mean size with  highly non-spherical surfaces due to $\Sigma(T, \mu) < 0$ and, hence, it is {\bf QGP in its traditional sense}. 

Besides the inequality $\tau > 2$ the Model 2   requires 
the fulfillment of several  additional conditions \cite{QGBSTM2,QGBSTM2b}. Thus, for   
$\mu \ge  \mu_{cep}$ the line  
$\Sigma(T, \mu) = 0$ coincides with the deconfinement PT line in the $\mu-T$-plane, i.e.
$T_{dec}(\mu) = T_{\Sigma}(\mu)$ for $\mu \ge  \mu_{cep}$, and  $T$-derivative of the reduced surface tension coefficient has a discontinuity at the deconfinement PT line, i.e.
$\frac{\partial \Sigma^+}{\partial T}  \neq 
\frac{\partial \Sigma^- }{\partial T} $ at $T = T_{dec}(\mu)$. 
In  this model the {\bf quark gluon liquid} can exist  inside the mixed phase only \cite{QGBSTM2}
whereas in the $\mu-T$-plane the QGP exists everywhere above  the line $\Sigma(T, \mu) = 0$. 
As a consequence,
the volume distribution of large bags of the Model 2  has the power law   right at the 
mixed phase \cite{QGBSTM2}. 
Obviously, the different location of the power law in volume distribution of large bags  
distinguishes models with the tricritical and with  critical endpoints and it can serve as
a clear experimental indicator to distinguish them. In  the Models 1 and 2 the volume of the QGP bag is proportional to its mass and, hence, one can also search for the power law 
in  the mass distribution of heavy bags \cite{Bugaev10}. 
This property remains also   valid within the more realistic statistical model which accounts for 
the finite width of the QGP bags \cite{FWM}. 
Therefore, the {\bf second challenge for  femtoscopy}
is {\it an elucidation of  the power law of the volume (mass)  distribution of large (heavy) bags from the available   data.}

It is necessary to stress  that the vast majority of statistical models is simply unrealistic since they do not account for the finite width of QGP bags. 
The latter is absolutely required  in order to naturally  explain \cite{FWM,Blascke:04} the huge existing   deficit in the number  of heavy hadronic resonances compared to the Hagedorn mass spectrum. 
Recently  within the finite width model 
\cite{FWM,FWMb}  it was shown  that even  in a vacuum  the mean  width of a resonance of mass $M$   behaves as $\Gamma (M) \approx 400-600 \left[\frac{M}{M_0}\right]^{\frac{1}{2}}$ MeV (with $M_0 \approx 2.5$ GeV), whereas in a media it increases  with the temperature.   
These results not only naturally explain the existing  deficit in the number  of heavy hadronic resonances mentioned above, but also allow us to establish a novel  view  at the confinement problem. Thus, usually the  confinement is understood as an  impossibility to separate 
the color charges confined by the gluonic fields. The finite width of large/heavy  QGP bags 
demonstrates another feature of  the confinement -- the large/heavy QGP bags are very unstable in the vacuum, i.e. they decay fast  without the stabilizing external conditions. 
Clearly, the large width of QGP bags should affect the space-time evolution of quark gluon matter (liquid or plasma)  created in the relativistic  nuclear collisions. 
Therefore, in my mind, {\bf the third challenge for femtoscopy \it is  an  investigation of the influence  of finite width of quark gluon bags on  their space-time evolution during the course 
of high energy nuclear collision.}
Some ideas on how to reach this goal are discussed in \cite{Testing:09}.


\medskip 

\noindent
{\bf 4. New challenges for femtoscopy.} In summary, the main challenges for femtoscopy are as follows:

\noindent
{\bf I. \it To study the emission from highly non-spherical (fractal) bags and  to 
find the indicator which is able to distinguish the case of positive, zero and negative surface tension coefficient.
}

\noindent
{\bf II. \it To elucidate  the power law of the volume (mass)  distribution of large (heavy) bags from the available  experimental  data.}

\noindent
{\bf III. \it To study  the influence  of finite width of quark gluon bags on  their space-time evolution during the course 
of high energy nuclear collision.}

\medskip

\textbf{ Acknowledgments.}  The research made in this work 
was supported   by the Program ``Fundamental Properties of Physical Systems 
under Extreme Conditions''  of the Bureau of the Section of Physics and Astronomy  of
the National Academy of Science of Ukraine.


\begin{thebibliography}{99}

\bibitem{Kink}
\refitem{article}
M. Gazdzicki, Z. Phys. {\bf C66}, 659 (1995);  J. Phys. {\bf G23}, 1881 (1997).

\bibitem{Horn}
\refitem{article}
M. Gazdzicki and M. I. Gorenstein, {Acta Phys. Polon.} {\bf B30}, 2705 (1999).

\bibitem{Step}
\refitem{article}
%
M. I. Gorenstein, M. Gazdzicki and K. A. Bugaev,
Phys. Lett. {\bf B567},  175 (2003).

\bibitem{Claims}
\refitem{article}
%
M. Gazdzicki, M. Gorenstein and P. Seyboth,  
arXiv:1006.1765 [hep-ph] (2010). 

\bibitem{QGBSTM}
\refitem{article}
%
K. A. Bugaev,
{Phys. Rev.} {\bf C 76},    014903 (2007).

\bibitem{QGBSTMb}
\refitem{article}
%
K. A. Bugaev,
{Phys. Atom. Nucl.} {\bf 71}, 1615  (2008).


\bibitem{QGBSTM2}
\refitem{preprint}
%
K. A. Bugaev, V. K. Petrov and G. M. Zinovjev,
arXiv:0904.4420  [hep-ph] (2009).

\bibitem{QGBSTM2b}
\refitem{article}
%
K. A. Bugaev, V. K. Petrov and G. M. Zinovjev, Europhys. Lett.
{\bf 85},  22002 (2009). 

\bibitem{String:09}
\refitem{article}
%
K. A. Bugaev  and G. M. Zinovjev,
Nucl. Phys. {\bf A 848}, 443  (2010). 

\bibitem{FWM}
\refitem{article}
%
%
K. A. Bugaev, V. K. Petrov and G. M. Zinovjev,
Phys. Rev. C {\bf   79},    054913 (2009). 

\bibitem{FWMb}
\refitem{article}
%
%
K. A. Bugaev and G. M. Zinovjev,
Ukr. J. Phys. {\bf 55},     586 (2010).



\bibitem{Jaffe:1}
%
\refitem{article}
E.~Farhi and R.~L.~Jaffe,   Phys. Rev. D \textbf{30},   2379 (1984).


\bibitem{Jaffe:2}
%
\refitem{article}
M.~S.~Berger and R.~L.~Jaffe,  Phys. Rev. C \textbf{35}, 213 (1987).



\bibitem{Fisher:67}
\refitem{article}
M.~E.~Fisher, Physics \textbf{ 3},  255 (1967).


\bibitem{Elliott:06}
%
\refitem{report}
J.~B.~Elliott, K.~A.~Bugaev, L.~G.~Moretto and L.~Phair,
arXiv:0608022 [nucl-ex]  (2006).

\bibitem{simpleSMM:1}
%
\refitem{article}
S.~Das~Gupta and A.~Z.~Mekjian, Phys. Rev. C \textbf{ 57}, 1361 (1998).

\bibitem{Landau:Statmech}
\refitem{book}
%
L. D. Landau and E.M. Lifshitz, {\it Statistical Physics}, (Fizmatlit, Moscow, 2001).

\bibitem{Shuryak:sQGP}
\refitem{article}
E. V. Shuryak,  
Prog. Part. Nucl. Phys. {\bf 62}, 48 (2009). 

\bibitem{Bugaev:04b}
%
\refitem{article}
K.~A.~Bugaev, L.~Phair and J.~B.~Elliott,
Phys. Rev. E \textbf{72},  047106  (2005).


\bibitem{Bugaev:05a}
%
\refitem{article}
K.~A.~Bugaev  and J.~B.~Elliott,
 Ukr. J. Phys. \textbf{52},  301 (2007).

\bibitem{StrTension2}
\refitem{article}
%
A. Mocsy  and P. Petreczky,
PoS LAT2007  {\bf 216} (2007).

%
\bibitem{Hagedorn:65}
\refitem{article}
%
R. Hagedorn, Nuovo Cimento Suppl. {\bf 3},  147 (1965).


\bibitem{Kapusta:81}
\refitem{article}
%
J. I.~Kapusta, Phys. Rev. D {\bf 23},  2444 (1981).


\bibitem{Bugaev10}
%
K. A. Bugaev,
arXiv:1012.3400  [nucl-th]  (2010).
  
\bibitem{Blascke:04}
%
D. B. Blaschke and K. A. Bugaev,
 Fizika B {\bf  13} (2004),  491.
  

\bibitem{Testing:09}
%
K. A. Bugaev, 
arXiv:0909.0731 [nucl-th]  (2009).



\end{thebibliography}
\end{document}